%% file: duotok.tex
\documentclass[10pt,twocolumn,letterpaper]{article}
\usepackage[margin=0.7in,columnsep=0.24in]{geometry}
\usepackage[T1]{fontenc}
\usepackage{newtxtext,newtxmath}
\usepackage{iftex}
\ifPDFTeX
  \usepackage[scaled=0.94]{helvet}
\else
\fi
\usepackage{graphicx,tabularx,enumitem,amsmath,booktabs,multirow,placeins,balance}
\usepackage{xcolor,titlesec,caption,tcolorbox,fancyhdr,needspace}
\usepackage[numbers,sort&compress]{natbib}
\usepackage{hyperref}
\definecolor{ink}{HTML}{1E3A8A}
\definecolor{brandblue}{HTML}{3B82F6}
\definecolor{labelgray}{HTML}{4C5259}
\definecolor{secondaryblue}{HTML}{60A5FA}
\definecolor{plotgreen}{HTML}{10B981}
\definecolor{baselinegray}{HTML}{9CA3AF}
\definecolor{plotbadge}{HTML}{DBEAFE}
\definecolor{gridgray}{HTML}{CBD5E1}
\definecolor{mist}{HTML}{F3F6FC}
\hypersetup{colorlinks=true,linkcolor=ink,citecolor=ink,urlcolor=ink,pdftitle={DuoTok: Source-Aware Dual-Track Music Tokenization for Vocal-Accompaniment Generation},pdfauthor={Rui Lin and Zhiyue Wu and Jiahe Lei and Kangdi Wang and Weixiong Chen and Junyu Dai and Tao Jiang}}
\titleformat{\section}{\large\sffamily\bfseries\color{black}}{\thesection}{0.6em}{}
\titleformat{\subsection}{\normalsize\sffamily\bfseries\color{black}}{\thesubsection}{0.5em}{}
\titlespacing*{\section}{0pt}{9pt plus 2pt minus 2pt}{4pt}
\titlespacing*{\subsection}{0pt}{7pt plus 2pt minus 1pt}{3pt}
\titleformat{\paragraph}[runin]{\normalfont\bfseries}{}{0pt}{}
\titlespacing*{\paragraph}{0pt}{5pt plus 1pt minus 1pt}{0.6em}
\pretocmd{\section}{\Needspace{5\baselineskip}}{}{}
\setlist[itemize]{leftmargin=*,nosep,topsep=3pt}
\providecommand{\Description}[1]{}
\input{sections/99_macros}

\begin{document}
\twocolumn[{
\begin{tcolorbox}[colback=mist,colframe=mist,boxrule=0pt,arc=5pt,left=12pt,right=12pt,top=10pt,bottom=10pt]
{\sffamily\bfseries\fontsize{21}{24}\selectfont\color{black}DuoTok: Source-Aware Dual-Track Music\\Tokenization for Vocal-Accompaniment Generation\par}
\vspace{7pt}
{\sffamily\fontsize{10.5}{14}\selectfont\bfseries Rui Lin\textsuperscript{*}\quad Zhiyue Wu\textsuperscript{*}\quad Jiahe Lei\quad Kangdi Wang\\Weixiong Chen\quad Junyu Dai\textsuperscript{\S\textdagger}\quad Tao Jiang\par}
\vspace{3pt}
{\sffamily\color{plotgreen}\ensuremath{\varepsilon}ar-lab\par}
\vspace{3pt}
{\footnotesize *Equal Contribution, \textsuperscript{\S}Corresponding Author, \textsuperscript{\textdagger}Project Lead\par}
{\footnotesize\sffamily \textcolor{black}{Contact:} \href{mailto:linr3639@gmail.com}{linr3639@gmail.com}\par}
\vspace{7pt}
{\sffamily\bfseries\small Abstract\par}\vspace{3pt}
{\small
Multi-track music generation requires token representations that are not only reconstructable, but also easy to model and structurally aligned across tracks. In the dual-track vocal--accompaniment setting, this requirement creates a fundamental tension: reconstruction-oriented codecs preserve acoustic detail but are difficult for sequence modeling, while more semantic tokenizers are often more predictable yet may lose fidelity or cross-track structure. We present \DuoTok{}, a source-aware dual-track music tokenizer designed as a multi-track generation interface through \emph{staged disentanglement}. \DuoTok{} first learns a semantic audio representation via self-supervised pretraining, then shapes source-aware structure in continuous space with feature replacement noise and multi-task supervision, including spectral reconstruction, music source separation regularization, and an ASR head for lyric alignment. It then freezes the encoder and learns hard-routed dual codebooks for vocals and accompaniment, while a diffusion decoder restores fine acoustic detail from the discrete tokens.

Across public benchmarks, \DuoTok{} achieves a better predictability--fidelity trade-off than prior tokenizers at ultra-low bitrate. Under held-constant dual-track language modeling, it further improves both unconditional vocal--accompaniment modeling and vocal-conditioned accompaniment prediction, indicating that its token space provides a stronger interface for dual-track sequence modeling. Controlled diagnostics show larger predictability costs under cross-track corruption and larger gains from longer temporal context, supporting the interpretation that these gains arise from stronger structural usage rather than merely easier local prediction. At the same time, \DuoTok{} maintains competitive reconstruction quality and preserves control-relevant musical attributes in the discrete space. Together, these results suggest that, for multi-track music generation, tokenizer design should be treated as a core modeling problem rather than only a compression problem.
\par}
\vspace{6pt}
{\scriptsize Author version of the paper accepted at ACM Multimedia 2026 (MM '26).\\
ACM publication DOI: \href{https://doi.org/10.1145/3767308.3836425}{10.1145/3767308.3836425}. \copyright\ 2026 the authors. Licensed under \href{https://creativecommons.org/licenses/by/4.0/}{CC BY 4.0}.\par}
\end{tcolorbox}
\vspace{5pt}
}]
\thispagestyle{plain}
\input{sections/01_intro}
\input{sections/02_related}
\input{sections/03_method}
\input{sections/04_experiments}
\input{sections/05_results}
\input{sections/06_analysis}
\input{sections/08_conclusion}
\FloatBarrier
\balance
\input{duotok.bbl}

\end{document}

%% file: sections/99_macros.tex
\providecommand{\DuoTok}{DuoTok}

\providecommand{\StageOne}{Stage~1}
\providecommand{\StageTwo}{Stage~2}
\providecommand{\StageThree}{Stage~3}
\providecommand{\StageFour}{Stage~4}

\providecommand{\cnBPT}{\mathrm{cnBPT}}
\providecommand{\cnBPS}{\mathrm{cnBPS}}
\providecommand{\enBPT}{\mathrm{enBPT}} 

\providecommand{\cnbpt}{\ensuremath{\cnBPT}}
\providecommand{\cnbps}{\ensuremath{\cnBPS}}
\providecommand{\enbpt}{\ensuremath{\enBPT}}

\providecommand{\figref}[1]{Fig.~\ref{#1}}
\providecommand{\tabref}[1]{Table~\ref{#1}}


%% file: sections/01_intro.tex
\section{Introduction}
\label{sec:intro}

Modern multi-track music generation increasingly depends on the tokenizer as the interface between waveforms and discrete sequence modeling~\citep{yue, levo, songcreator}. In the dual-track setting, vocals and accompaniment are modeled as synchronized but distinct streams, enabling conditional generation, track-level editing, and finer control than mixture-only modeling. In this setting, the tokenizer is not merely a compression front-end: it serves as a \emph{multi-track generation interface} that directly shapes what the model can reconstruct, predict, and coordinate across tracks.

However, existing tokenizers remain poorly matched to this setting. Reconstruction-oriented codecs such as EnCodec and DAC preserve rich acoustic detail, but their token sequences retain abundant fine-grained variability and are difficult to model autoregressively~\citep{encodec, dac, wavtokenizer}. Semantic-oriented tokenizers move toward more predictable representations, but may weaken the timbre, texture, and detailed fidelity needed for high-quality music reconstruction~\citep{semanticcodec, speechtokenizer, xcodec}. More fundamentally, in the multi-track setting, cross-track correspondence is rarely treated as a first-class constraint in the token space. What is still missing, therefore, is not simply another codec, but a tokenizer that can jointly support predictability, reconstruction, and cross-track structure.

\begin{figure}[t]
    \centering
    \includegraphics[width=0.9\linewidth]{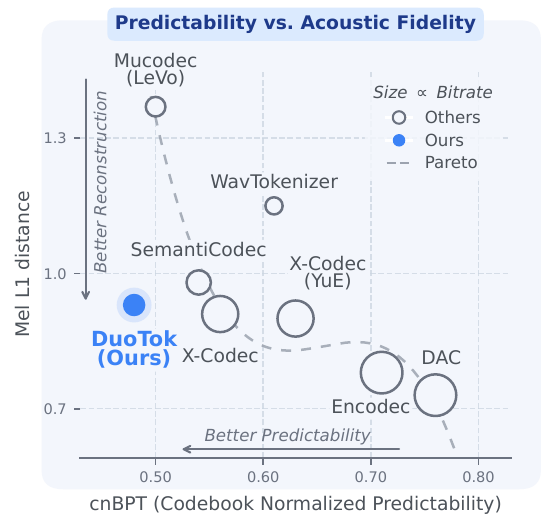}
    \caption{\textbf{Predictability--fidelity trade-off across tokenizers.}
    Bubble size denotes bitrate. Lower \cnbpt{} and lower Mel indicate a better trade-off.}
    \Description{A Pareto-style scatter plot comparing tokenizers on normalized predictability (cnBPT) and reconstruction Mel distance. Each point is a tokenizer, and bubble size indicates bitrate. Methods closer to the lower-left region achieve a better predictability--fidelity trade-off.}
    \label{fig:pareto}
\end{figure}

We address this problem with \DuoTok{}, a source-aware dual-track music tokenizer built on \emph{staged disentanglement}. The core idea is to first shape source-aware structure in continuous space, and only then discretize it into dual-track token streams. This ordering avoids forcing the quantizer to operate on under-structured representations, and prevents the discrete token space from carrying the full burden of micro-acoustic detail. Instead, the model first forms a representation that is better structured, easier to model, and more aware of source correspondence, before applying hard routing for track-specific discretization.

Concretely, \DuoTok{} is trained in four stages. \StageOne{} builds a semantic foundation through self-supervised pretraining. \StageTwo{} shapes source-aware structure with feature replacement noise and auxiliary supervision, including spectral reconstruction, music source separation regularization, and an ASR head for lyric alignment. \StageThree{} freezes the encoder and learns hard-routed dual codebooks for vocals and accompaniment. \StageFour{} decouples rendering from token learning through diffusion-based reconstruction, restoring fine acoustic detail while allowing the discrete interface to focus on structured information for sequence modeling.

This design yields a better balance for dual-track music modeling. As shown in Fig.~\ref{fig:pareto}, \DuoTok{} improves the predictability--fidelity trade-off over prior tokenizers. Under held-constant dual-track LM evaluation, it also improves both unconditional vocal--accompaniment modeling and vocal-conditioned accompaniment prediction, indicating a more modelable and structurally useful token space for dual-track sequence modeling. Controlled diagnostics further show larger penalties under cross-track corruption and larger gains from longer context, indicating that these gains are tied to stronger structural usage rather than merely easier local-token prediction. At the same time, controlled reconstruction probes and tagging results show that \DuoTok{} remains reconstructable and preserves control-relevant musical attributes even at low bitrate.

Our contributions are threefold:
\begin{itemize}
    \item \textbf{A new tokenizer formulation for multi-track generation.} We treat music tokenization as a source-aware dual-track generation interface rather than only a compression module.
    
    \item \textbf{A staged tokenizer design.} We propose \DuoTok{}, which uses staged disentanglement to separate semantic formation, source-aware structure shaping, stable dual-track discretization, and high-fidelity rendering.
    
    \item \textbf{A held-constant evaluation and evidence chain for tokenizer quality.} We show that \DuoTok{} improves the predictability--fidelity trade-off, strengthens unconditional and conditional dual-track LM modeling under matched settings, and is further supported by structural diagnostics, controlled reconstruction probes, and ablations.
\end{itemize}

%% file: sections/02_related.tex
\section{Related Work}
\label{sec:related}

\subsection{Audio Tokenizers for Generation}

Discrete audio tokenizers define the token sequences used for downstream sequence modeling, so tokenizer design directly shapes the trade-off between reconstruction fidelity and sequence modelability. Reconstruction-oriented neural codecs, such as SoundStream, EnCodec, DAC, and WavTokenizer, prioritize perceptual quality and compression efficiency, but their acoustic tokens retain substantial fine-grained variability and are therefore harder to model over long contexts~\citep{soundstream, encodec, dac, wavtokenizer}.

Another line of work derives semantic tokens from self-supervised audio representations by clustering, quantizing, or otherwise discretizing continuous SSL features. This includes speech-oriented SSL foundations as well as music-aware representation models such as MERT, BEATs, and MuQ~\citep{wav2vec2, hubert, wavlm, bestrq, mert, beats, muq}. Such tokens are often more LM-friendly, but can be too lossy for music detail. Recent hybrid tokenizers therefore inject SSL priors into codec-style discretization to better balance predictability and fidelity~\citep{semanticcodec, speechtokenizer, xcodec, xytokenizer}.

A closely related direction more directly brings LM objectives or LM-side pressures into tokenizer training, in order to preserve intelligibility, modelability, or text alignment under reconstruction-oriented settings, including recent systems such as Ming UniAudio, MOSS audio tokenizer, and MIMO audio tokenizer~\citep{ming_uniaudio, moss_audio_tokenizer, mimo}. These methods make tokenizer learning more LM-aware or text-aligned, but their primary focus is not the design of a synchronized dual-track interface for vocal--accompaniment modeling.

Still, most existing tokenizers remain single-stream or mixture-centric, and rarely treat cross-track correspondence as a tokenizer-level objective. In contrast, our focus is not only to make tokens more intelligible or LM-friendly in isolation, but to design a source-aware dual-track interface that jointly balances predictability, reconstructability, and cross-track correspondence.

\subsection{Multi-Track Music Generation}

Token-based music generation commonly follows a tokenizer--LM--renderer pipeline. While earlier large-scale systems mainly adopt single-stream modeling, recent work increasingly moves toward structured multi-track generation. In particular, dual-track systems such as YuE, SongCreator, and LeVo explicitly separate vocals and accompaniment to improve controllability, editability, and conditional generation~\citep{musiclm, songgen, yue, songcreator, levo}. In this setting, the tokenizer is no longer only a codec front-end, but a multi-track generation interface that shapes how synchronized token sequences can be modeled and coordinated.

Existing multi-track tokenization strategies, however, remain limited. Shared or mixture-centric token spaces can entangle vocals and accompaniment, while separation-first approaches often treat tracks as largely independent and leave cross-track structure to the sequence model. As a result, current tokenizers rarely encode source separability and cross-track correspondence jointly, leaving a gap for a dual-track tokenizer that supports reconstructability, LM-friendly sequence modeling, and structure-aware vocal--accompaniment modeling.

%% file: sections/03_method.tex
\section{\DuoTok{}}
\label{sec:method}

\paragraph{Design goals.}
\DuoTok{} is a source-aware dual-track tokenizer designed as a \emph{multi-track generation interface} for vocal--accompaniment music modeling.
Its design is guided by three goals:
\textbf{predictability}, so the resulting token streams are easy to model under multi-track LMs;
\textbf{semantic preservation}, so the representation retains track-aware structure and control-relevant musical attributes;
and \textbf{reconstruction fidelity}, so the discrete interface remains decodable with sufficient acoustic detail.
To balance these goals, \DuoTok{} uses \emph{staged disentanglement}: it first builds a generic semantic foundation in continuous space, then shapes it into a music-domain representation that is more source-aware and text-aligned, then discretizes it into dual-track token streams, and finally decouples high-fidelity rendering from token learning.

\subsection{Overview}
\label{ssec:overview}

\begin{figure*}[t]
  \centering
  \includegraphics[width=0.9\textwidth]{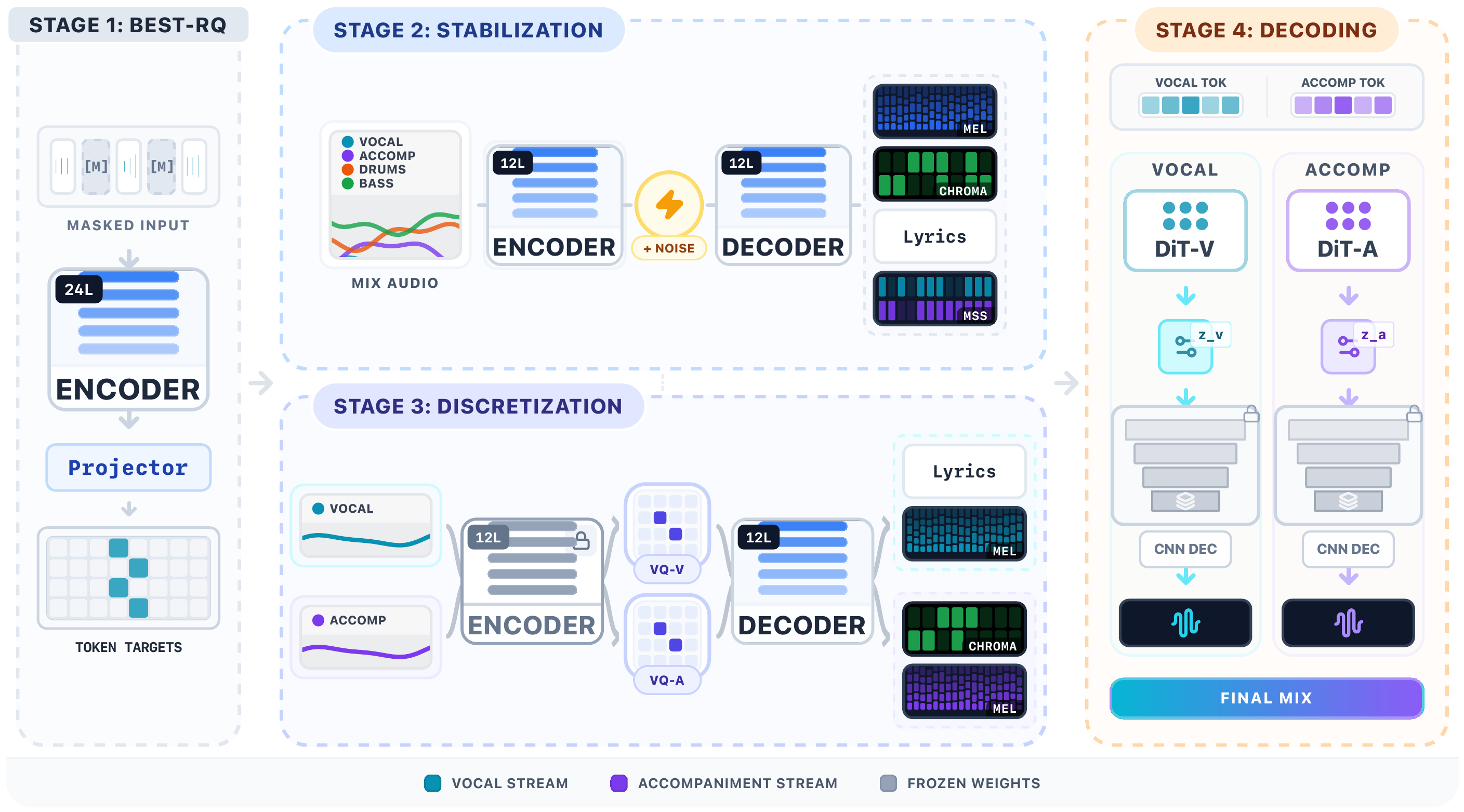}
  \caption{\textbf{Overview of \DuoTok{}.}
  \StageOne{} learns a semantic foundation. \StageTwo{} shapes music-domain, source-aware, and lyric-aligned structure in continuous space. \StageThree{} freezes the encoder and discretizes features into hard-routed dual-track codebooks. \StageFour{} reconstructs waveforms with diffusion decoders.}
  \Description{A four-stage pipeline diagram of DuoTok from left to right. Stage 1 performs semantic self-supervised pretraining. Stage 2 adds source-aware and lyric-aligned auxiliary objectives with feature replacement noise. Stage 3 freezes the encoder and applies hard-routed dual codebooks for vocal and accompaniment tokenization. Stage 4 uses diffusion decoders to reconstruct high-fidelity waveforms from discrete tokens.}
  \label{fig:pipeline}
\end{figure*}

\DuoTok{} is trained in four stages, as shown in Fig.~\ref{fig:pipeline}.
\StageOne{} learns a semantic audio representation through self-supervised pretraining.
\StageTwo{} adapts this representation to the music domain while shaping source-aware and lyric-aligned structure through feature replacement noise and auxiliary supervision.
\StageThree{} freezes the encoder and discretizes the learned representation into vocal and accompaniment streams with hard-routed dual codebooks.
\StageFour{} freezes the tokenizer and trains diffusion decoders to restore fine acoustic detail.
This ordering lets the token space focus on structured information for sequence modeling, instead of carrying the full burden of waveform rendering.

\subsection{Continuous-space structure formation}
\label{ssec:continuous}

\paragraph{\StageOne{}: semantic foundation.}
\StageOne{} learns a high-level audio representation with self-supervised pretraining~\citep{wav2vec2,hubert,wavlm}.
We adopt BEST-RQ~\citep{bestrq} to provide a semantic foundation that is more content-oriented and better suited for later tokenization.
This stage is intentionally generic and single-stream: rather than directly optimizing reconstruction, it establishes a semantically grounded and more globally learnable representation that later stages can adapt to music structure.

\paragraph{\StageTwo{}: source-aware structure shaping.}
Starting from the pretrained encoder, \StageTwo{} trains the encoder together with auxiliary prediction heads to adapt the representation toward the music domain while preserving the semantic regularity learned in \StageOne{}.
The goal is not simply to add acoustic detail, but to shape the continuous representation so that it becomes more source-aware, more lyric-aligned, and more suitable for later discretization.
We inject feature replacement noise by randomly replacing a subset of hidden states with Gaussian noise:
\begin{align}
\tilde{\mathbf{h}}_t &= (1-m_t)\mathbf{h}_t + m_t \boldsymbol{\epsilon}_t, \\
\boldsymbol{\epsilon}_t &\sim \mathcal{N}\!\left(\mathbf{0},\,\sigma^2\mathbf{I}\right),
\end{align}
where $m_t$ is a Bernoulli mask.
This noise injection discourages the representation from overfitting to fragile local acoustic cues and helps prevent it from collapsing into low-level acoustic patterns, thereby preserving the global learnability established during pretraining.
We then optimize
\begin{equation}
\label{eq:stage2_loss}
\mathcal{L}^{(2)}
=
\mathcal{L}_{\mathrm{mel}}
+\lambda_{\mathrm{chr}}\mathcal{L}_{\mathrm{chr}}
+\lambda_{\mathrm{asr}}\mathcal{L}_{\mathrm{asr}}
+\lambda_{\mathrm{mss}}\mathcal{L}_{\mathrm{mss}}.
\end{equation}
Here $\mathcal{L}_{\mathrm{mel}}$ and $\mathcal{L}_{\mathrm{chr}}$ provide spectral reconstruction supervision, $\mathcal{L}_{\mathrm{asr}}$ is applied through an \emph{ASR head for lyric alignment}, and $\mathcal{L}_{\mathrm{mss}}$ distills source-aware structure from a frozen MSS teacher via mask prediction over $N_{\mathrm{src}}$ teacher sources:
\begin{equation}
\mathcal{L}_{\mathrm{mss}} = \sum_{s=1}^{N_{\mathrm{src}}}\left\lVert \hat{M}^{(s)} - M^{(s)} \right\rVert_{1}.
\end{equation}
Together, these objectives make the continuous representation not only reconstructable, but also more explicitly organized around source correspondence and lyric alignment.
The purpose of \StageTwo{} is to shape a continuous representation that is already semantically grounded, music-adapted, source-aware, and stable enough for hard dual-track discretization.

\subsection{Stable dual-track discretization and rendering decoupling}
\label{ssec:discrete}

\paragraph{\StageThree{}: frozen-encoder dual-track discretization.}
In \StageThree{}, we freeze the encoder learned from \StageOne{}--\StageTwo{} and train only the discretization module and the heads attached to the quantized representation.
This design preserves the structured continuous representation before hard routing, rather than asking the quantizer to create cross-track structure from scratch.
Given a track indicator $r \in \{\text{vocal}, \text{accompaniment}\}$, features are routed to a track-specific quantizer with codebook $\mathcal{C}_r$.
We adopt a linear-reparameterized SimVQ~\citep{simvq} as a more stable parameterization for hard dual-track discretization, especially in the large-codebook setting.
We then quantize in the transformed space $\tilde{\mathcal{C}}_r = \mathcal{C}_r W_r$:
\begin{equation}
k_t = \arg\min_{j} \|\mathbf{h}_t - \tilde{\mathbf{c}}^{(r)}_j\|_2^2,
\qquad
\mathbf{q}_t = \tilde{\mathbf{c}}^{(r)}_{k_t}.
\end{equation}
This hard routing produces separate but time-aligned token streams for vocals and accompaniment.
During this stage, we keep the Mel, Chroma, and lyric-alignment heads on the quantized representation, and jointly optimize
\begin{equation}
\label{eq:stage3_loss}
\mathcal{L}^{(3)}
=
\mathcal{L}_{\mathrm{mel}}
+\lambda_{\mathrm{chr}}\mathcal{L}_{\mathrm{chr}}
+\lambda_{\mathrm{asr}}\mathcal{L}_{\mathrm{asr}}
+\lambda_{\mathrm{vq}}\mathcal{L}_{\mathrm{vq}}.
\end{equation}
Since \StageThree{} operates on track-separated pseudo-stems, MSS supervision is not used in this stage.
The purpose of \StageThree{} is to discretize an already structured representation into a stable dual-track interface.

\paragraph{\StageFour{}: rendering decoupling.}
In \StageFour{}, we freeze the tokenizer and train track-wise latent diffusion decoders conditioned on \StageThree{} tokens.
These decoders predict pre-trained ear-VAE latents~\citep{earvae} and reconstruct the final waveform.
This stage decouples high-fidelity rendering from token learning, so the discrete interface can focus on structured information that is more useful for multi-track sequence modeling rather than micro-acoustic detail.
In our current instantiation, we use track-wise latent diffusion decoders for high-fidelity rendering; however, because rendering is decoupled from token learning, the learned dual-track token interface is not conceptually tied to this specific decoder choice.

\paragraph{Summary.}
Overall, \DuoTok{} first builds a generic semantic foundation, then shapes it into a music-domain continuous representation that is source-aware and lyric-aligned, then freezes and discretizes that representation into a stable dual-track token interface, and finally restores fine acoustic detail with a separate renderer.
This staged design is intended to improve the predictability--fidelity trade-off while better supporting cross-track sequence modeling.

%% file: sections/04_experiments.tex
\section{Experimental Setup}
\label{sec:exp}

\subsection{Data, pseudo-stems, and supervision}
\label{ssec:data}

\DuoTok{} is trained exclusively on public corpora.
\StageOne{} uses a broad mixture of speech, general audio, and music to learn a semantic audio representation.
\StageTwo{}--\StageFour{} focus on music and vocal-centric data.

To enable dual-track learning without proprietary multi-track recordings, we construct pseudo vocal and accompaniment stems with Demucs~\citep{demucs}.
Audio is segmented into 5--30\,s clips, and when lyrics are available, we apply lyric-aware segmentation aligned to sentence boundaries.

Because pseudo-stem pipelines can introduce separator-specific artifacts, we deliberately decouple the two sources of source-related supervision used in \DuoTok{}.
In \StageTwo{}, MSS regularization is distilled from a frozen BS-RoFormer teacher~\citep{bsroformer}, which predicts soft time--frequency masks for MUSDB-style sources.
This supervision is used only as a weak structural prior in continuous space: we do not directly regress separated waveforms in this stage.
In \StageThree{}, by contrast, the tokenizer is trained on track-separated pseudo vocal/accompaniment stems produced by Demucs.
These two components therefore serve different roles and are not the same source: the former provides continuous-space structural regularization, while the latter provides the dual-track inputs used for discretization.

For lyric supervision, we attach an ASR head for lyric alignment.
Chinese and English lyric transcripts are obtained with WhisperX and FunASR~\citep{whisperx,funasr}.
Across stages, we optimize with AdamW and a warmup--cosine schedule; full data mixtures and hyperparameters are provided in the supplement.

\subsection{Baselines}
\label{ssec:baselines}

We compare against representative tokenizers from three families:
\textbf{reconstruction-oriented} codecs, including DAC, EnCodec, and WavTokenizer;
\textbf{semantic-oriented} tokenizers, including SemantiCodec and X-Codec;
and \textbf{music-oriented} tokenizers used in recent multi-track generation systems, including X-Codec in YuE and MuCodec in LeVo.
We use official checkpoints and standard configurations whenever available.

\subsection{Held-constant evaluation protocols}
\label{ssec:protocols}

We evaluate tokenizers with standardized, held-constant protocols.

\paragraph{Multi-track LM evaluation.}
Across tokenizers, we train the same \texttt{Qwen2-1B} backbone under the same optimizer, learning-rate schedule, total optimization steps, and 30\,s audio context.
We evaluate both \textbf{unconditional dual-track modeling} and \textbf{track-conditioned modeling}, where the target track is predicted given the other track.
This protocol is intended to isolate tokenizer quality for dual-track sequence modeling rather than confounding it with different LM backbones or training budgets.

\paragraph{Semantic accessibility probe.}
To test whether the discrete space preserves control-relevant musical attributes, we freeze each tokenizer and train a lightweight probe using only post-quantization codebook embeddings.
We evaluate on MagnaTagATune Top-50, whose tags cover common control-relevant attributes such as genre, instrumentation, mood, and vocal-related cues, and report AP and AUC.

\paragraph{Reconstruction evaluation.}
We report each tokenizer's native reconstruction pipeline to reflect end-to-end system behavior.
To reduce decoder confounds, we additionally perform a matched-decoder comparison by training the same latent diffusion decoder family on frozen codebook embeddings to predict ear-VAE latents under matched training settings. This controlled comparison is intended to probe the reconstructability of the token space under fixed decoder capacity, rather than to claim a perfectly decoder-neutral benchmark.
We further validate perceptual quality with a blinded MOS study using randomized presentation and report mean $\pm$ standard error.

\subsection{Metrics}
\label{ssec:metrics}

\paragraph{Predictability.}
We measure token-level cross-entropy in bits per token (BPT).
Because raw BPT scales with codebook size, our main unified metric is codebook-normalized BPT:
\begin{equation}
\mathrm{cnBPT}
=
\frac{\mathrm{BPT}}
{\frac{1}{K_{\mathrm{cb}}}\sum_{k=1}^{K_{\mathrm{cb}}}\log_2 S_k},
\label{eq:cnbpt_main}
\end{equation}
where $K_{\mathrm{cb}}$ is the number of codebooks and $S_k$ is the size of codebook $k$.
Lower cnBPT indicates better predictability after a first-order correction for codebook size, so we use it as the primary comparison metric across tokenizers.

When the empirical code distribution is reliably estimable, we additionally report entropy-normalized BPT:
\begin{equation}
\mathrm{enBPT}=\frac{\mathrm{BPT}}{H(p)},
\label{eq:enbpt_main}
\end{equation}
where $H(p)$ is the empirical unigram entropy on the evaluation set.
Since consistent entropy estimation is not available for all compared tokenizers, enBPT is used as a complementary metric in the held-constant dual-track LM benchmark.

Because cnBPT is a per-token quantity, tokenizers with similar cnBPT can still impose different per-second modeling burdens if they emit different numbers of tokens per second.
We therefore additionally report codebook-normalized bits per second:
\begin{equation}
\mathrm{cnBPS}=\tau\cdot\mathrm{cnBPT},
\label{eq:cnbps_main}
\end{equation}
where $\tau$ is the total token emission rate in tokens/s, summed over streams when applicable.
We use cnBPS as a rate-aware complementary metric for cross-tokenizer trade-off comparisons, while keeping cnBPT as the primary unified predictability metric.

Taken together, cnBPT, enBPT, and cnBPS play complementary roles: cnBPT provides the main unified comparison across tokenizers, while enBPT and cnBPS help verify whether the same conclusions remain consistent under entropy-aware and rate-aware normalization.

\paragraph{Semantic accessibility.}
We report Average Precision (AP) and Area Under the ROC Curve (AUC) on MagnaTagATune Top-50.

\paragraph{Reconstruction.}
We use Mel distance as the main reconstruction metric, together with objective perceptual metrics where appropriate.
In matched-decoder evaluation, we additionally report Mean Opinion Score (MOS) to assess perceptual quality.

Overall, we first evaluate the predictability--fidelity trade-off across tokenizers, and then test whether the resulting token space provides a stronger interface for dual-track language modeling under held-constant evaluation.

%% file: sections/05_results.tex
\section{Main Results}
\label{sec:results}

We first establish the overall predictability--fidelity trade-off, then test whether the tokenizer provides a stronger \emph{dual-track sequence-modeling interface} under held-constant dual-track language modeling, and finally examine whether these gains come at the expense of reconstruction or control-relevant musical attributes.

\subsection{Overall Predictability--Fidelity Trade-off}
\label{ssec:overall_tradeoff}

\figref{fig:pareto} summarizes the trade-off landscape between predictability and acoustic fidelity. Reconstruction-oriented codecs~\citep{encodec,dac,wavtokenizer} remain strong in waveform recovery, but tend to produce token sequences that are harder to model. Semantic-oriented tokenizers~\citep{semanticcodec,xcodec} move toward lower modeling complexity, but often at the cost of reconstruction quality. For dual-track music modeling, however, a useful tokenizer should not simply optimize one side of this trade-off: it should provide a better balance between sequence modelability and reconstruction quality.

Within this landscape, \DuoTok{} offers the most favorable overall balance among the compared methods. As shown in \tabref{tab:main_results}, it achieves the lowest \cnbpt{} at 0.75\,kbps while maintaining competitive reconstruction quality. The comparison with MuCodec (LeVo) is especially important because the two models operate at closely matched bitrates: \DuoTok{} improves \cnbpt{} from 0.50 to 0.48 while also reducing Mel distance from 1.37 to 0.95. It also achieves the strongest tagging AP at this bitrate. Taken together, these results indicate that \DuoTok{} improves the predictability--fidelity trade-off rather than gaining LM-friendliness by collapsing fidelity.

Because tokenizers also differ in token emission rate, \tabref{tab:main_results} additionally reports \cnbps{} as a rate-aware complementary metric. The same conclusion remains consistent under this per-second view: \DuoTok{} slightly improves over the closely matched MuCodec baseline (24.0 vs.\ 25.0) while maintaining a clear reconstruction advantage. This suggests that the gain is not simply a per-token normalization artifact.

Overall trade-off performance is strong, but this does not yet show whether \DuoTok{} provides a stronger token space for dual-track language modeling; next we test that directly.

\begin{table*}[t]
    \centering
    \scriptsize
    \caption{\textbf{Overall benchmark comparison.}
    AP/AUC, predictability (\cnbpt{}, \cnbps{}), and reconstruction (Mel). Metric directions follow table headers.}
    \label{tab:main_results}
    \setlength{\tabcolsep}{4pt}
    \begin{tabular}{@{}lcccccccc@{}}
        \toprule
        & \multicolumn{3}{c}{Metadata} & \multicolumn{2}{c}{Tagging} & \multicolumn{2}{c}{Predictability} & \multicolumn{1}{c}{Recon.} \\
        \cmidrule(lr){2-4} \cmidrule(lr){5-6} \cmidrule(lr){7-8} \cmidrule(l){9-9}
        Model & Rate & Size ($K_{\mathrm{cb}}{\times}S$) & Kbps & MTT-AP $\uparrow$ & AUC $\uparrow$ & \textbf{cnBPT} $\downarrow$ & \textbf{cnBPS} $\downarrow$ & Mel $\downarrow$ \\
        \midrule
        \textit{Reconstruction-oriented} \\
        DAC \cite{dac}                    & 75 Hz  & $8{\times}1024$   & 6.00 & 0.20 & 0.79 & 0.76 & 456.0 & \textbf{0.73} \\
        EnCodec \cite{encodec}            & 75 Hz  & $8{\times}1024$   & 6.00 & 0.18 & 0.76 & 0.71 & 426.0 & 0.78 \\
        WavTokenizer \cite{wavtokenizer}  & 40 Hz  & $1{\times}4096$   & 0.48 & 0.17 & 0.74 & 0.61 & 24.4  & 1.15 \\
        \midrule
        \textit{Semantic-oriented} \\
        SemantiCodec \cite{semanticcodec} & 50 Hz  & $2{\times}8192$   & 1.30 & 0.32 & \textbf{0.88} & 0.54 & 54.0  & 0.98 \\
        X-Codec \cite{xcodec}             & 50 Hz  & $8{\times}1024$   & 4.00 & 0.32 & 0.87 & 0.56 & 224.0 & 0.91 \\
        \midrule
        \textit{Music-oriented} \\
        X-Codec (YuE) \cite{yue}            & 50 Hz & $16{\times}1024$  & 8.00 & 0.32 & 0.87 & 0.63 & 504.0 & 0.90 \\
        MuCodec (LeVo) \cite{levo,mucodec}  & 25 Hz & $2{\times}16384$  & 0.70 & 0.26 & 0.84 & 0.50 & 25.0  & 1.37 \\
        \textbf{\DuoTok{} (Ours)}           & 25 Hz & $2{\times}32768$  & 0.75 & \textbf{0.35} & 0.87 & \textbf{0.48} & \textbf{24.0} & 0.95 \\
        \bottomrule
    \end{tabular}
\end{table*}

\subsection{Dual-Track Sequence-Modeling Ability}
\label{ssec:multitrack_generation}

We now evaluate the tokenizer as a dual-track modeling interface under the held-constant dual-track LM protocol. Here, \cnbpt{} is the primary comparison metric, while \enbpt{} and Acc@\(k\) are reported as supporting diagnostics. The key question is not only whether the token sequences are easy to model in isolation, but whether they support synchronized modeling across vocal and accompaniment streams.

\textbf{Unconditional dual-track modeling.}
As shown in \tabref{tab:uncond_results}, \DuoTok{} achieves the lowest average \cnbpt{} (\textbf{0.483}), improving over X-Codec in YuE (0.631) and MuCodec in LeVo (0.501). The same overall trend remains under \enbpt{}, where \DuoTok{} matches the strongest baseline and substantially outperforms X-Codec in YuE. Although its larger vocabulary makes raw top-\(k\) accuracy less suitable as a primary cross-model metric, \DuoTok{} remains competitive in Acc@\(k\), indicating that the gains in normalized BPT are not accompanied by obvious degradation in supporting accuracy diagnostics. We also report tokenizer-level codebook-usage diagnostics in this table: \DuoTok{} maintains near-full utilization on both codebooks (0.9996/0.9995) with high normalized entropy (0.9188/0.9494), indicating healthy large-codebook occupancy rather than partial collapse.

\textbf{Vocal-conditioned accompaniment prediction.}
The advantage is even clearer in track-conditioned modeling. In \tabref{tab:cond_results}, \DuoTok{} achieves the lowest predictability cost under both normalization schemes, reaching \textbf{0.464} \cnbpt{} and \textbf{0.489} \enbpt{}. This task more directly probes whether the vocal token stream exposes useful structure for predicting the accompaniment stream. Its advantage here is therefore especially important: it suggests that \DuoTok{} is not merely easier to model within each track, but exposes a more useful synchronized interface for cross-track prediction.

Taken together, the unconditional and conditioned results consistently support a more modelable and structurally useful token space for dual-track sequence modeling, rather than a gain confined to one track or one evaluation direction. These gains raise a deeper question: do they reflect stronger structural usage, or merely easier local prediction? We address this question in Section~\ref{sec:why_it_works}{} through structural diagnostics and design ablations.

\begin{table}[t]
    \centering
    \scriptsize
    \caption{\textbf{Unconditional dual-track LM under the held-constant protocol.}
    Primary metric: \cnbpt{}; \enbpt{}, Acc@\(k\), and codebook usage are supporting diagnostics. Util. is used-entry fraction; $H_{\mathrm{norm}}$ is normalized unigram entropy.}
    \label{tab:uncond_results}
    \setlength{\tabcolsep}{0.8pt}
    \renewcommand{\arraystretch}{0.95}
    \begin{tabular}{@{}llc ccc cccc@{}}
        \toprule
        \multirow{2}{*}{Model} & \multirow{2}{*}{Size} & \multirow{2}{*}{Track}
        & \multicolumn{3}{c}{Acc}
        & \multirow{2}{*}{cnBPT $\downarrow$} & \multirow{2}{*}{enBPT $\downarrow$}
        & \multirow{2}{*}{Util. $\uparrow$} & \multirow{2}{*}{$H_{\mathrm{norm}}$ $\uparrow$} \\
        \cmidrule(lr){4-6}
        & & & @1 & @10 & @50 & & & & \\
        \midrule

        \multirow{3}{*}{X-Codec (YuE)} & \multirow{3}{*}{$8{\times}1024$}
            & Vocal  & \textbf{0.15} & 0.44 & \textbf{0.70} & 0.632 & 0.773 & 0.9162 & 0.8165 \\
        & & Accomp & \textbf{0.15} & \textbf{0.45} & \textbf{0.73} & 0.630 & 0.735 & 0.9299 & 0.8579 \\
        & & Avg    & --   & --   & --   & 0.631 & 0.754 & -- & -- \\
        \midrule

        \multirow{3}{*}{MuCodec (LeVo)} & \multirow{3}{*}{$2{\times}16384$}
            & Vocal  & \textbf{0.15} & \textbf{0.45} & 0.67 & 0.485 & \textbf{0.497} & 0.9929 & \textbf{0.9767} \\
        & & Accomp & \textbf{0.15} & 0.41 & 0.64 & 0.517 & 0.536 & 0.9719 & \textbf{0.9660} \\
        & & Avg    & --   & --   & --   & 0.501 & \textbf{0.516} & -- & -- \\
        \midrule

        \multirow{3}{*}{\textbf{\DuoTok{}}} & \multirow{3}{*}{$2{\times}32768$}
            & Vocal  & 0.14 & 0.43 & 0.67 & \textbf{0.461} & 0.501 & \textbf{0.9996} & 0.9188 \\
        & & Accomp & 0.11 & 0.36 & 0.61 & \textbf{0.506} & \textbf{0.533} & \textbf{0.9995} & 0.9494 \\
        & & Avg    & --   & --   & --   & \textbf{0.483} & \textbf{0.516} & -- & -- \\
        \bottomrule
    \end{tabular}
\end{table}

\begin{table}[t]
    \centering
    \scriptsize
    \caption{\textbf{Vocal-conditioned accompaniment prediction under the held-constant protocol.}
    Primary metric: \cnbpt{}; \enbpt{} and Acc@\(k\) are supporting diagnostics.}
    \label{tab:cond_results}
    \setlength{\tabcolsep}{1pt}
    \begin{tabular}{@{}lccccccc@{}}
        \toprule
        \multirow{2}{*}{Model} & \multirow{2}{*}{Size}
        & \multicolumn{3}{c}{Acc}
        & \multirow{2}{*}{cnBPT $\downarrow$} & \multirow{2}{*}{enBPT $\downarrow$} \\
        \cmidrule(lr){3-5}
        & & @1 & @10 & @50 & & \\
        \midrule

        X-Codec (YuE)      & $8{\times}1024$  & 0.15 & \textbf{0.46} & \textbf{0.73} & 0.593 & 0.691 \\
        MuCodec (LeVo)     & $2{\times}16384$ & \textbf{0.17} & \textbf{0.46} & 0.69 & 0.488 & 0.505 \\
        \textbf{\DuoTok{}} & $2{\times}32768$ & 0.13 & 0.41 & 0.66 & \textbf{0.464} & \textbf{0.489} \\
        \bottomrule
    \end{tabular}
\end{table}

\subsection{Reconstruction and Semantic Accessibility}
\label{ssec:recon_semantic}

A stronger tokenizer should not be one-dimensional. If lower predictability were obtained only by discarding recoverable detail or useful control-related information, the resulting token space would be less valuable in practice. We therefore examine whether \DuoTok{} preserves both reconstructability and control-relevant musical attributes.

\textbf{Why matched-decoder evaluation matters.}
Raw reconstruction comparisons across tokenizers can be confounded by decoder design. We therefore complement native-pipeline reconstruction results with a matched-decoder comparison between \DuoTok{} and MuCodec (LeVo), using the same latent diffusion decoder family conditioned on frozen codebook embeddings. This controlled setting is intended to reduce decoder confounds and probe the reconstructability of the token space under fixed decoder capacity, although it does not fully eliminate decoder--token compatibility effects.

\textbf{Controlled reconstruction.}
As shown in \tabref{tab:controlled_recon}\textbf{(a)}, \DuoTok{} improves vocal reconstruction across all reported objective metrics, including PESQ, STOI, and Mel distance. For accompaniment, where speech-oriented metrics are less appropriate, \DuoTok{} also achieves a substantially lower Mel distance. The same pattern carries into perceptual evaluation: the blinded MOS results in \tabref{tab:controlled_recon}\textbf{(b)} show a clear subjective preference for \DuoTok{} on both vocal and accompaniment tracks. These results support the view that the LM-facing gains are not simply purchased by a collapse in recoverability.

\textbf{Semantic accessibility.}
We further evaluate whether the discrete representation preserves control-relevant musical attributes using the MTT tagging benchmark. In \tabref{tab:main_results}, a probe trained on frozen post-quantization embeddings reaches an AP of \textbf{0.35} and an AUC of 0.87 for \DuoTok{}, giving the strongest AP among the compared methods at an ultra-low bitrate. We interpret this result as evidence of \emph{semantic accessibility}: the token space retains information useful for downstream control-related attribute prediction. We do not treat this as a claim of broad music understanding, but as support that improved predictability is not achieved by discarding control-relevant information.

Overall, \DuoTok{} is not a one-dimensional LM gain: it achieves a stronger balance across predictability, reconstructability, and control-relevant semantic accessibility.

\begin{table}[t]
    \centering
    \small
    \caption{\textbf{Controlled reconstruction under the matched-decoder protocol.}
    \textbf{(a)} Objective metrics. \textbf{(b)} Blinded MOS. Metric directions follow table headers.}
    \label{tab:controlled_recon}
    \setlength{\tabcolsep}{5pt}

    \begin{tabular}{@{}llccc@{}}
        \toprule
        \multicolumn{5}{@{}l}{\textit{(a) Objective metrics}} \\
        Type & Codec & PESQ $\uparrow$ & STOI $\uparrow$ & Mel $\downarrow$ \\
        \midrule
        Vocal  & MuCodec (LeVo)     & 1.28 & 0.35 & 1.39 \\
               & \textbf{\DuoTok{}} & \textbf{1.76} & \textbf{0.55} & \textbf{0.81} \\
        \midrule
        Accomp & MuCodec (LeVo)     & --   & --   & 1.85 \\
               & \textbf{\DuoTok{}} & --   & --   & \textbf{1.14} \\
        \bottomrule
    \end{tabular}

    \vspace{3pt}

    \begin{tabular}{@{}lccc@{}}
        \toprule
        \multicolumn{4}{@{}l}{\textit{(b) Subjective MOS (1--5)}} \\
        Type & GT & MuCodec (LeVo) & \textbf{\DuoTok{}} \\
        \midrule
        Vocal  & 4.61 $\pm$ 0.12 & 1.51 $\pm$ 0.09 & \textbf{3.66 $\pm$ 0.05} \\
        Accomp & 4.52 $\pm$ 0.19 & 1.85 $\pm$ 0.11 & \textbf{3.07 $\pm$ 0.12} \\
        \bottomrule
    \end{tabular}
\end{table}

%% file: sections/06_analysis.tex
\section{Why It Works: Structural Evidence and Robustness}
\label{sec:why_it_works}

Section~\ref{ssec:multitrack_generation} shows that \DuoTok{} provides a more modelable and structurally useful token space for dual-track sequence modeling. We now ask why these gains arise. In particular, are they driven by stronger use of cross-track and temporal structure, or merely by easier local token prediction? A second concern is whether the observed gains could be overstated by pseudo-stem artifacts or a particular separator pipeline. We address these questions through structural diagnostics, compact robustness checks, and design ablations.

\subsection{Structural Diagnostics and Robustness}
\label{ssec:structural_diagnostics}

We test whether the dual-track LM gains in Section~\ref{ssec:multitrack_generation} are genuinely structural. If \DuoTok{} mainly simplifies local token statistics, then breaking cross-track correspondence or extending temporal context should have limited effect. If it instead exposes stronger source correspondence and non-local structure, correspondence-breaking perturbations should incur larger penalties and longer context should yield larger gains. Because these analyses rely on pseudo-stem inputs, we also check robustness under separator swap and controlled cross-track contamination.

\begin{figure}[t]
    \centering
    \includegraphics[width=0.8\linewidth]{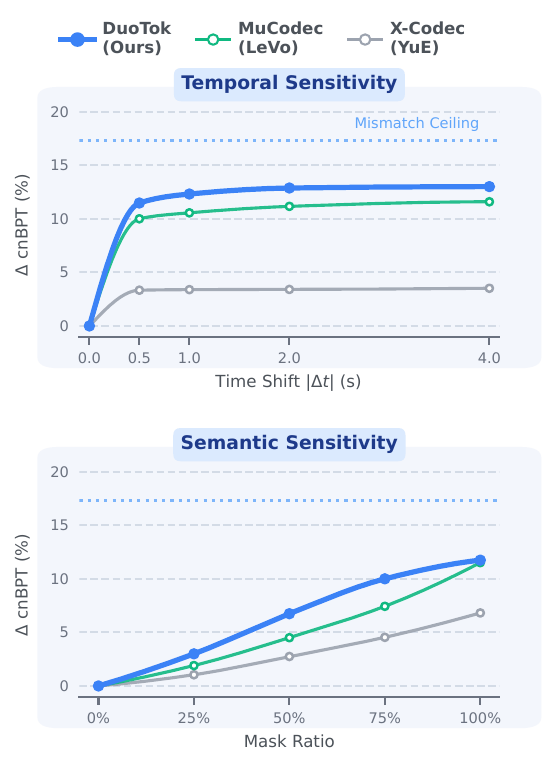}
    \caption{\textbf{Cross-track corruption sensitivity.}
    Relative \cnbpt{} increase under temporal shift, masking, and uncorrelated mismatch.}
    \Description{A multi-curve comparison of relative cnBPT increase under three conditioning-track corruptions: temporal shift, random masking ratio, and uncorrelated mismatch. DuoTok shows larger degradation than weaker baselines under correspondence-breaking perturbations, indicating stronger use of cross-track structure.}
    \label{fig:sensitivity}
\end{figure}

To quantify the effect of correspondence-breaking perturbations, we report the relative predictability cost under corruption:
\begin{equation}
\Delta \mathrm{cnBPT}(\%) =
100 \cdot
\frac{\mathrm{cnBPT}_{\mathrm{corrupt}} - \mathrm{cnBPT}_{\mathrm{clean}}}
{\mathrm{cnBPT}_{\mathrm{clean}}}.
\label{eq:delta_cnbpt}
\end{equation}
We consider three perturbation types on the conditioning track: \textbf{temporal shift}, \textbf{random masking}, and \textbf{uncorrelated mismatch}, where the conditioning input is replaced by an unrelated sample. As shown in Fig.~\ref{fig:sensitivity}, all perturbations increase predictability cost, and mismatch is consistently the most damaging. \DuoTok{} shows stronger degradation under correspondence-breaking corruption than weaker baselines, while shift and masking also produce clear monotonic penalties. Together, these trends indicate stronger reliance on cross-track correspondence rather than only local token regularities.

\begin{figure}[t]
    \centering
    \includegraphics[width=0.8\linewidth]{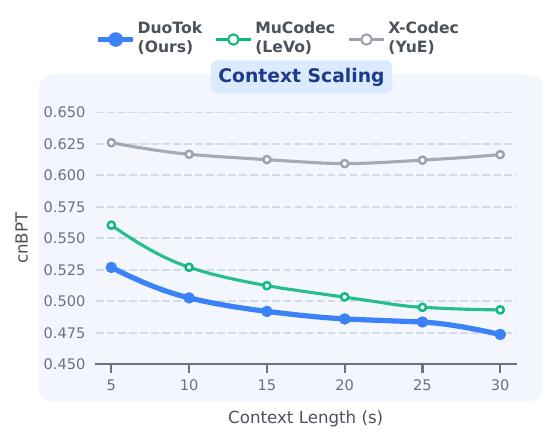}
    \caption{\textbf{Context scaling under the held-constant dual-track LM protocol.}
    \cnbpt{} versus context length (lower is better).}
    \Description{A line chart of cnBPT versus context length for compared tokenizers under the same dual-track language model setup. cnBPT decreases as context grows, and DuoTok has the steepest downward trend, showing better long-context utilization.}
    \label{fig:context_scaling}
\end{figure}

We next test temporal-context utilization. Figure~\ref{fig:context_scaling} plots \cnbpt{} under increasing context length. Both \DuoTok{} and MuCodec (LeVo) benefit from longer context, but \DuoTok{} shows a steeper reduction, indicating stronger context-scaling efficiency. This suggests the gains in Section~\ref{ssec:multitrack_generation} are not limited to easier short-range prediction.

Because the dual-track setup relies on pseudo-stems, we test robustness to separator choice and controlled leakage. Table~\ref{tab:robustness_block} consolidates both checks. In \textbf{(a)} (separator swap), replacing the original separator with Demucs MDX-Net Extra produces only a small upward shift in \cnbpt{} while preserving the same monotonic context-length trend. In \textbf{(b)} (controlled contamination), leakage increases \cnbpt{} for both tokenizers, while \DuoTok{} is less sensitive at low contamination levels. These checks do not remove pseudo-stem concerns, but reduce the likelihood that the main findings are artifacts of a single separator pipeline or trivial cross-track leakage.

\begin{table}[t]
    \centering
    \small
    \caption{\textbf{Compact robustness block for pseudo-stem sensitivity.}
    \textbf{(a)} Separator swap. \textbf{(b)} Controlled cross-track contamination. Values in parentheses are relative increases from clean.}
    \label{tab:robustness_block}
    \setlength{\tabcolsep}{2.8pt}

    \resizebox{\linewidth}{!}{%
\begin{tabular}{@{}lcccccc@{}}
        \toprule
        \multicolumn{7}{@{}l}{\textit{(a) Separator swap under context scaling}} \\
        Separator & 5\,s & 10\,s & 15\,s & 20\,s & 25\,s & 30\,s \\
        \midrule
        Original & 0.5268 & 0.5026 & 0.4919 & 0.4859 & 0.4835 & 0.4736 \\
        Demucs MDX-Net Extra & 0.5400 & 0.5150 & 0.5040 & 0.4970 & 0.4930 & 0.4850 \\
        \bottomrule
    \end{tabular}
}

    \vspace{2pt}

    \resizebox{\linewidth}{!}{%
\begin{tabular}{@{}lcccc@{}}
        \toprule
        \multicolumn{5}{@{}l}{\textit{(b) Cross-track contamination robustness}} \\
        Tokenizer & 0\% & 1\% & 5\% & 10\% \\
        \midrule
        DuoTok & 0.4736 & 0.4780 (+0.93\%) & 0.4950 (+4.52\%) & 0.5186 (+9.50\%) \\
        MuCodec & 0.4931 & 0.5170 (+4.85\%) & 0.5340 (+8.29\%) & 0.5500 (+11.54\%) \\
        \bottomrule
    \end{tabular}
}
\end{table}

Taken together, \DuoTok{}'s gains are best interpreted as structural gains rather than simple token-level simplification, and the robustness checks make it less likely that this interpretation is dominated by a single pseudo-stem or separator pipeline.

\subsection{Design Ablations}
\label{ssec:design_ablations}

Having established a structural interpretation that is reasonably robust to separator choice and controlled leakage, we next ask whether the observed gains arise from the staged design itself or from incidental implementation details.

\begin{table}[t]
    \centering
    \small
    \caption{\textbf{Ablation summary for key Stage 2--3 design choices.}
    Unconditional LM predictability and reconstruction for the full model and ablated variants.}
    \label{tab:main_ablation}
    \setlength{\tabcolsep}{2.2pt}
    \renewcommand{\arraystretch}{0.95}
    \begin{tabular}{@{}lccccc@{}}
        \toprule
        Tokenizer
        & \shortstack{Vocal\\cnBPT $\downarrow$}
        & \shortstack{Accomp\\cnBPT $\downarrow$}
        & \shortstack{Avg.\\cnBPT $\downarrow$}
        & \shortstack{Vocal\\Mel $\downarrow$}
        & \shortstack{Accomp\\Mel $\downarrow$} \\
        \midrule
        \DuoTok{}
        & 0.4607 & \textbf{0.5057} & 0.4832 & \textbf{0.240} & 0.219 \\
        \quad w/o \textsc{MSS}
        & \textbf{0.4469} & 0.5129 & \textbf{0.4800} & 0.249 & 0.223 \\
        \quad w/o Noise
        & 0.5148 & 0.5599 & 0.5374 & 0.248 & 0.219 \\
        \quad w/o Enc Freeze
        & 0.6079 & 0.5929 & 0.6004 & 0.250 & \textbf{0.210} \\
        \bottomrule
    \end{tabular}
\end{table}

Removing feature replacement noise substantially hurts modelability: Avg.\ \cnbpt{} rises from 0.4832 to 0.5374, with both tracks worsening (vocal: 0.4607 \(\rightarrow\) 0.5148; accompaniment: 0.5057 \(\rightarrow\) 0.5599). In contrast, Mel changes are small (vocal: 0.240 \(\rightarrow\) 0.248; accompaniment: 0.219 \(\rightarrow\) 0.219). This pattern supports the intended role of noise as a structural regularizer for LM-friendly representations rather than a direct reconstruction booster.

Removing encoder freezing causes the largest degradation (Avg.\ \cnbpt{} 0.6004), including sharp increases on both tracks (vocal: 0.6079; accompaniment: 0.5929). Mel behavior is mixed: vocal Mel worsens to 0.250, while accompaniment Mel improves to 0.210. This indicates that once hard dual-codebook routing is introduced, keeping the encoder fixed is critical for stable, LM-friendly discretization.

The \textsc{MSS} ablation reveals a nuanced fidelity--predictability trade-off: removing \textsc{MSS} slightly improves Avg.\ \cnbpt{} (0.4832 \(\rightarrow\) 0.4800) but worsens Mel on both tracks, especially vocals (0.240 \(\rightarrow\) 0.249). This is consistent with \textsc{MSS} preserving source-aware spectral detail that is not always the easiest for the LM to model.

Taken together, the ablations support staged disentanglement as a design strategy rather than a single isolated trick: noise and encoder freezing are central to modelability, while \textsc{MSS} mainly improves fidelity at a small predictability cost.

%% file: sections/08_conclusion.tex
\section{Conclusion}
\label{sec:conclusion}

We frame music tokenization as a multi-track generation interface between waveforms and discrete sequence modeling. \DuoTok{} combines semantic pretraining, source-aware structure shaping, hard dual-codebook routing, and diffusion decoding to jointly support reconstruction, control-relevant attributes, predictable tokens, and cross-track correspondence.

Across benchmarks, \DuoTok{} achieves a strong predictability--fidelity balance, with the lowest \cnbpt{} and strongest tagging AP among compared methods while maintaining competitive reconstruction at ultra-low bitrate. Held-constant dual-track LM evaluation and structural diagnostics are consistent with stronger use of cross-track and temporal structure rather than only easier local prediction. These results position multi-track tokenizer design as an interface problem, not only a compression problem.